# A theory of web traffic

M.V. Simkin and V.P. Roychowdhury
*Department of Electrical Engineering, University of California, Los Angeles, CA 90095-1594 and NetSeer Inc. - 11943 Montana Ave. Suite 200, Los Angeles, CA 90049*

**Abstract –** We analyze access statistics of several popular webpages for a period of several years. The graphs of daily downloads are highly non-homogenous with long periods of low activity interrupted by bursts of heavy traffic. These bursts are due to avalanches of blog entries, referring to the page. We quantitatively explain this behavior using the theory of branching processes. We extrapolate these findings to construct a model of the entire web. According to the model, the competition between webpages for viewers pushes the web into a self-organized critical state. In this regime, the most interesting webpages are in a near-critical state, with a power-law distribution of traffic intensity.

The science of Self Organized Criticality (SOC) [1] was in the center of attention for the last two decades because it can explain power law distributions and bursts of intermittent activity, observed in many natural and social phenomena. An apparent downside of the SOC models is that they are heuristic. Examples include a model of interacting species on a circle for biological evolution, a block-spring model for earthquakes, and the Bak-Tang-Wiesenfeld (BTW) [2] sandpile model for everything. In the present letter, we introduce a new SOC model, which is not heuristic, but appears to describe what is really happening. Our model deals with web traffic.

Figure 1 shows access statistics for three popular webpages. Access statistics for half a dozen other popular webpages that we studied look very similar. The traffic is non-uniform in time with periods of low activity interrupted by bursts. To understand what is going on, we need to know where the visitors came from. This we can do by studying the referral statistics. Figure 2 shows the graphs of daily referrals to the webpage, whose access statistics is shown in Figure 1(a), from four different websites. Plots (a) and (b) of Fig. 2 are referrals from Google search and from Mozilla directory. They show a low intensity, but constant, stream of referrals. The distribution of the number of daily referrals follows a Poisson law. We will refer to such referrers as *constant referrers*. In plots (c) and (d) of Fig. 2, which show referrals from blogs, we see bursts of referrals. These bursts start with a peak (at the time when the link is blogged) and afterward subside, as the blog entry gets older. We will refer to such referrers as *temporary referrers*.

Now we can understand Fig. 1. The webpage in question gets traffic from constant referrers, which are responsible for all of the traffic between bursts. They play similar role to that of the falling grains in the BTW [2] sandpile model. Sometimes a visitor, who visited the webpage, following a constant referrer, creates a blog entry or makes a forum posting which links to the webpage. A reader of this blog, or forum, in his turn can link to the webpage in his own blog. If the webpage is interesting to many people, it can trigger avalanches of blog and forum postings. These avalanches are responsible for the bursts seen in Fig.1. Brain [3] and Arbesman [4] already discussed this mechanism of web traffic dynamics, but did not develop any mathematical theory. Here we construct a mathematical model of web traffic using the theory of branching processes [5].

Galton and Watson invented the theory of branching processes [5] in 1875, to explain the extinction of prominent British families. They considered a model where in each generation, $p(0)$ percent of the adult males have no sons, $p(1)$ have one son and so on. Using the theory one can compute the probability distribution of the sizes of families after any number of generations. If we graphically represent the family history by connecting each individual with his sons, we get a tree-like structure. This is where the name "branching process" comes from. The fate of families depends on the average number of sons $\lambda = \sum np(n)$. When $\lambda < 1$, the branching process is subcritical, that is all families eventually get extinct. When $\lambda > 1$, the branching process is supercritical, and some families do survive (and those, which survive, grow exponentially).

The branching process with $\lambda = 1$ is called critical. All families eventually become extinct, but the distribution of the lifetimes of families follows a power law. The distribution of combined offspring (the sum of the numbers of suns, grandsons and so on) follows a power law with exponent 1.5. In the case of a subcritical branching process, the distribution of combined offspring follows the same power law, only with an exponential cutoff. This cutoff becomes less and less short as $\lambda$ increases, eventually resulting in a pure power law when $\lambda$ reaches the value of 1. In the case of a supercritical branching process, combined offspring of a final fraction of families is equal to infinity[1]. Alstrøm [6] had shown that the mean-field version the BTW model is equivalent to a critical branching process. The combined offspring corresponds to the size of an avalanche in SOC. Note, that SOC is not merely a re-invention of the branching process, as SOC models have a *built in mechanism* for tuning the branching process into critical state.

[1] The distribution of combined offspring should not be confused with the distribution of subtrees. In Ref. [10] it was reported that the latter distribution is a power law with exponent 2 for a supercritical process. The distribution of subtrees is defined for a family tree after a large but fixed number of generations. Here one looks at the probability distribution of the sizes of subtrees rooted at a randomly chosen site. In the case of a subcritical or critical process the distribution of subtrees is equivalent to the distribution of combined offspring because when the number of generations is sufficiently large every subtree has time to terminate due to family extinction. The case of a supercritical process is different: considering only a finite number of generations cuts many would be infinite subtrees .

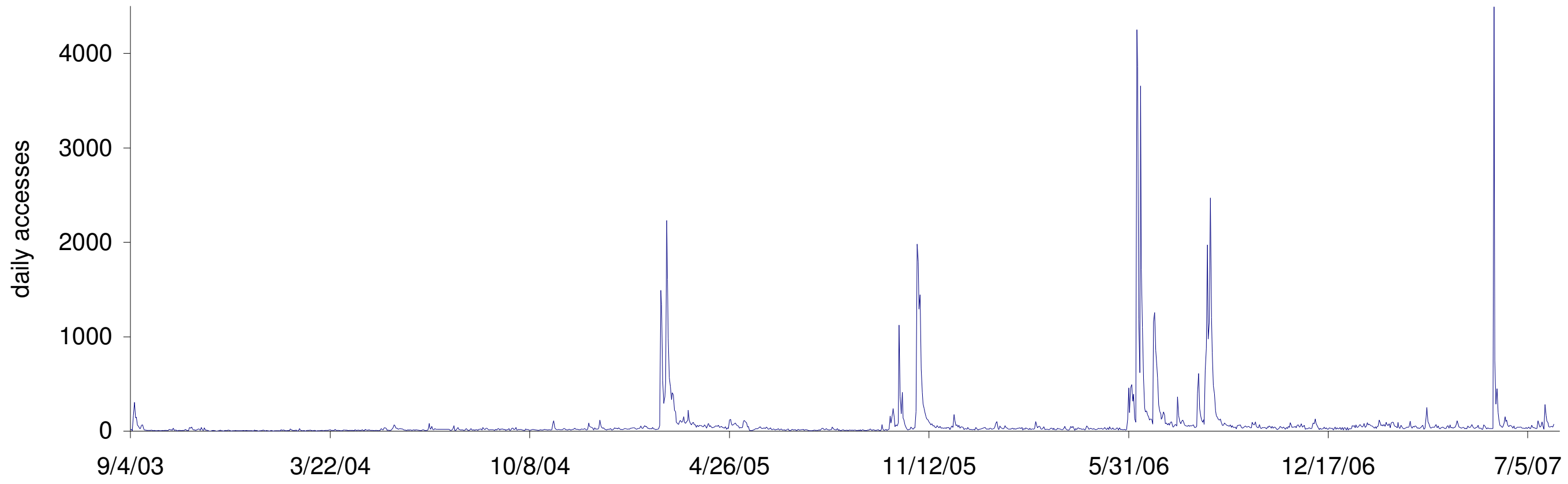

(a)

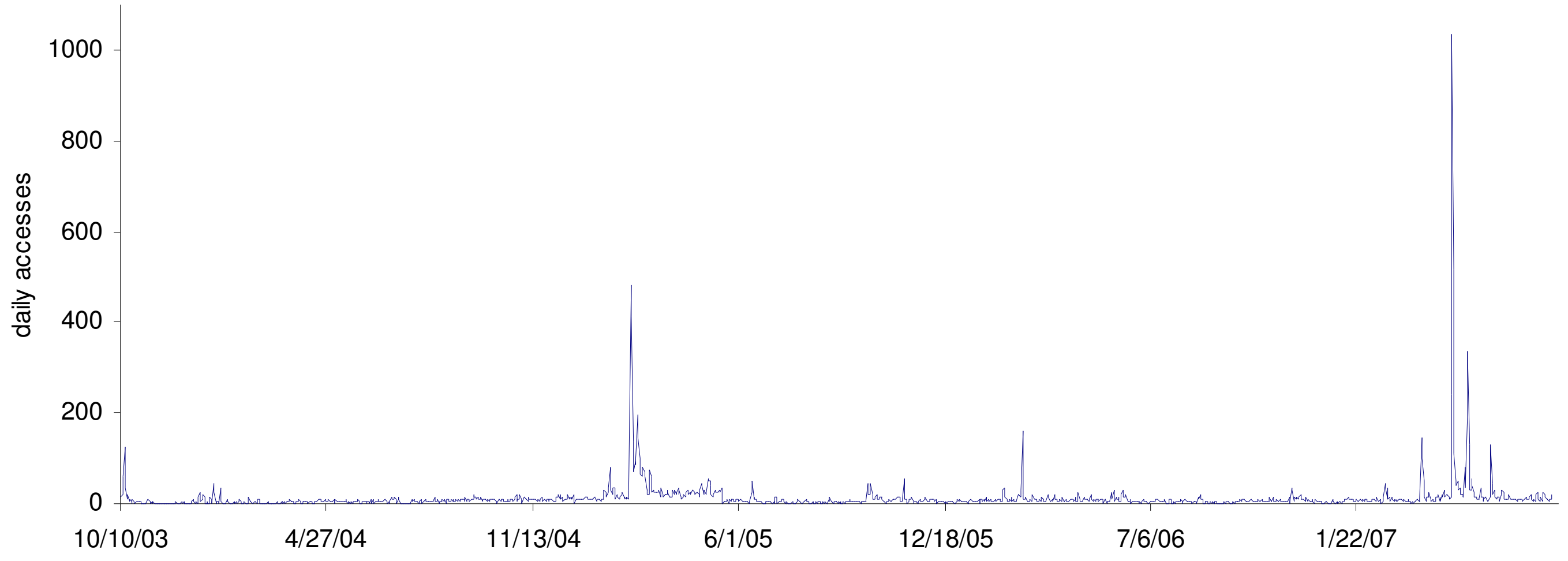

(b)

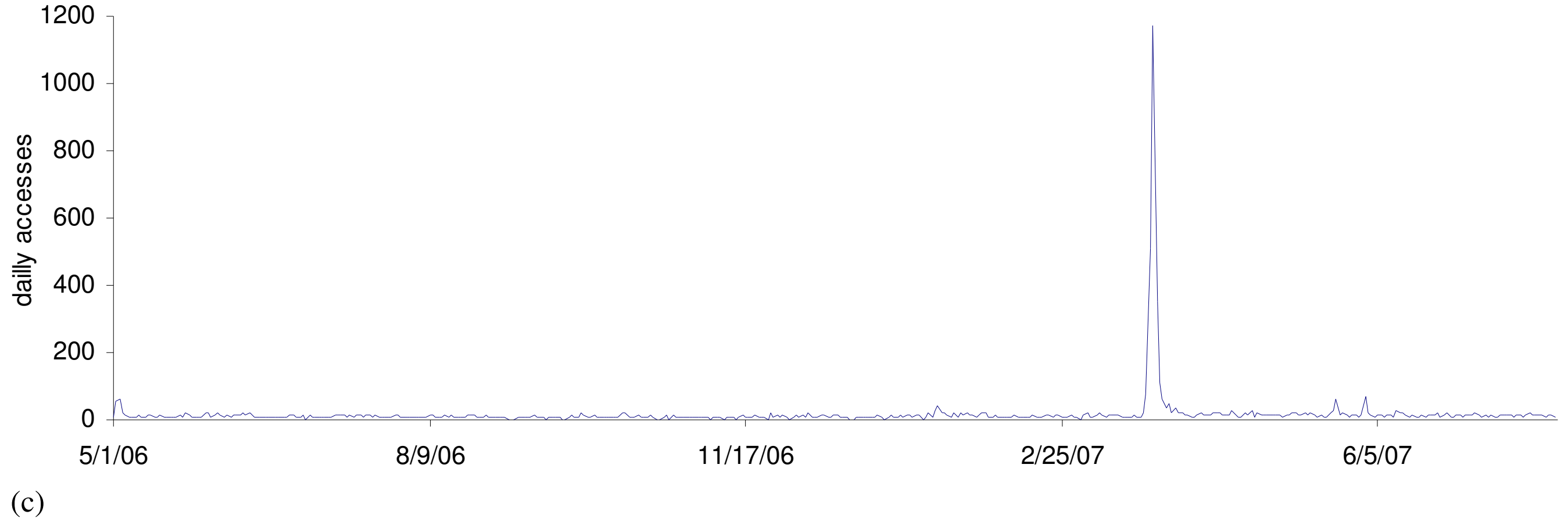

(c)

**Figure 1.** Access statistics for three webpages
(a) http://reverent.org/true_art_or_fake_art.html
(b) http://reverent.org/sounds_like_faulkner.html
(c) http://ecclesiastes911.net/disumbrated_art.html
starting from their creation dates and continuing until July 31 2007. Access statistic of a dozen other popular webpages, that we studied, looks very similar.

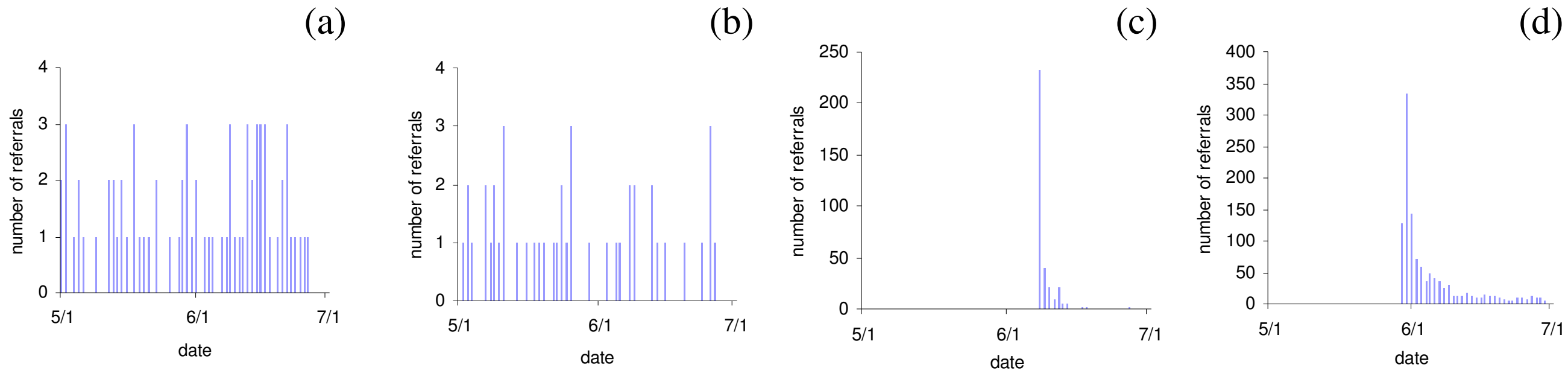

**Figure 2**. Statistics of referrals during two month in 2006 to the webpage, whose access statistics is shown in Fig. 1 (a), from four different websites**:**
(a) Google search; (b) Mozilla directory; (c) Presurfer - http://presurfer.meepzorp.com ;
(d) Reality Carnival - http://sprott.physics.wisc.edu/pickover/pc/realitycarnival.html .

To make our model tractable we introduce time-discretization with a unit of one day. The traffic from constant referrers we model as a Poisson process, with the number of daily referrals from all constant referrers following a Poisson distribution with mean $N_c$. To account for the fact that nobody reads old blog entries we assume that people read only today's posts. In reality people do read entries that are few days or weeks old, as it is evidenced by Fig. 2 (c) and (d). It is straightforward to incorporate this into the model, but it will only make it more complicated without adding much understanding. Thus, we assume that traffic from temporary referrers lives for one day. We assume that each visitor to the webpage will on the next day link to it in a temporary referrer with the probability $r$. We assume that each link in a temporary referrer generates, on average, $N$ visitors to our webpage. If we start with a single visitor and there are no additional referrals from constant referrers - we have a standard branching process. In the simplest case, when all temporary referrers refer exactly $N$ visitors, the offspring probabilities are $p(N) = r$, $p(0) = 1 - r$ and $p(n) = 0$ for the rest of $n$. As long as $\lambda = \sum np(n) = rN < 1$, this branching process is subcritical and the traffic must always eventually stop. However, thanks to constant referrers, many branching processes start every day: every visitor, referred by a constant referrer, starts a separate branching process.

The above model is easy to simulate on a computer. Figure 3 shows the graph of the number of daily downloads produced by such simulation. Although theory of branching processes gives quantitative predictions for the distribution of the sizes of avalanches, we cannot compare these predictions with the data of Figure 1, because avalanches overlap and there is no way to separate web accesses resulting from different avalanches. However, one can compare Zipf's plots of simulated and actual daily and monthly downloads. They are shown in Figure 4. The parameters used in the simulation were: $N_c = 10$, $r = 0.01$, and $N = 95$ (we assumed an exponential distribution of the number of visitors from a temporary referrer). We experimented with different values of parameters, but the above appear to produce the most similar outcome to the actual access statistics shown in Fig.1. With our parameters we get $\lambda = rN = 0.95$, which means that the branching process is slightly subcritical.

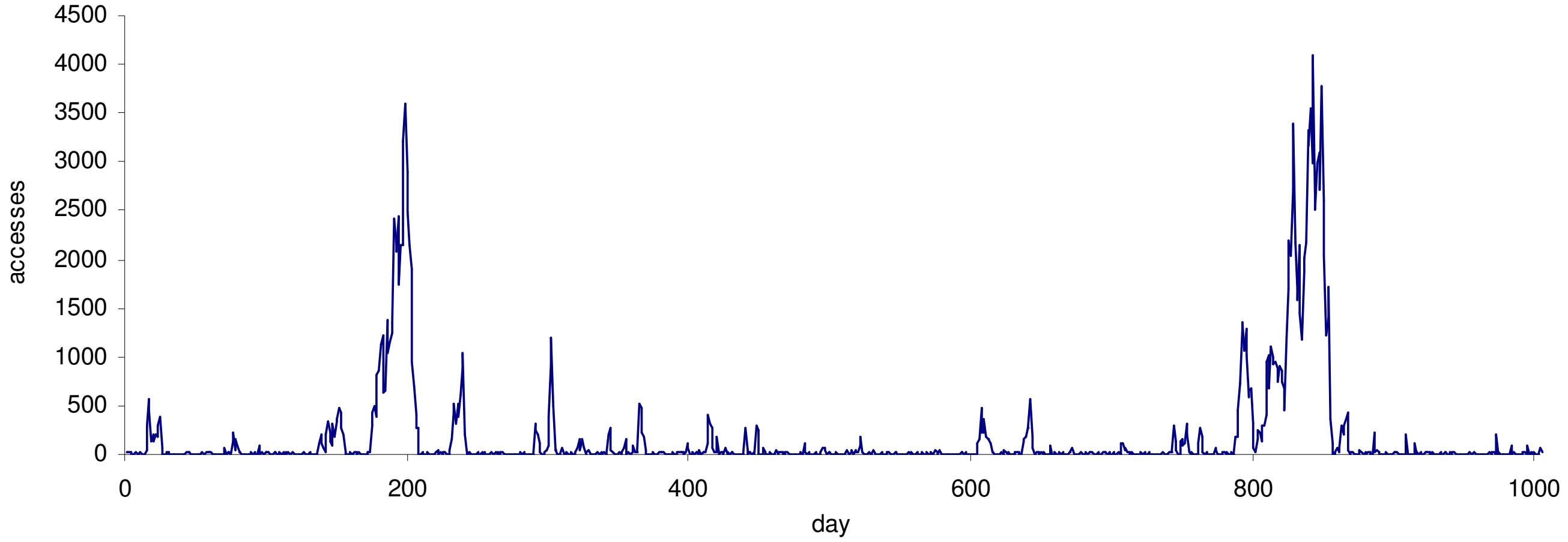

**Figure 3.** The outcome of one numerical simulation of the branching web traffic model.

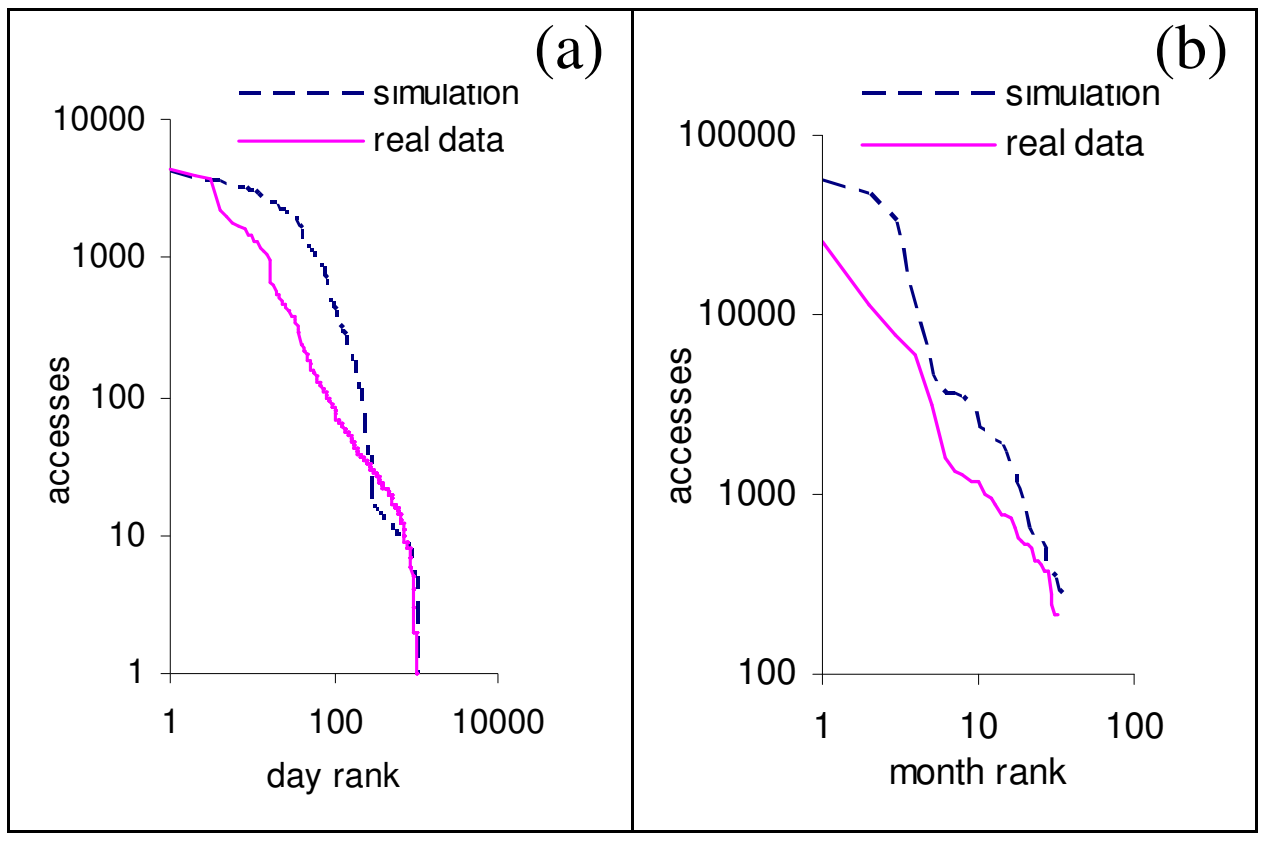

**Figure 4.** Zipf's plot of daily (a) and monthly (b) downloads computed using the data of Fig. 1(a) and Fig. 3.

It is interesting to compare Fig. 4 (b) with Nielsen's [7] Zipf's plot of the number downloads of many different webpages (from one website) during one month. Our plot, which shows the distribution of daily accesses of a single webpage during many months, looks very similar. This suggests that a part of variance in webpage popularity, observed in [4], is due to pure chance (variation in monthly webpage popularity). However, not all of the variance in popularity is due to chance. Notice that in Fig. 4(b) even the worst month had hundreds of accesses, while the number of downloads of the least popular webpage in Ref. [7] is one. In addition, most webpages whose access statistics we studied show low activity, uninterrupted by any bursts. This happens because apart from chance, an important factor in webpage popularity is its fitness, or ability to "resonate" [3].

Now we will use the understanding gained from studying one webpage access statistics, to formulate a model of the dynamics of the entire web. Recently we used a similar model to develop the mathematical theory of citing [8].

Let us consider a blogosphere in which $N_l$ blog entries are added daily. The WWW apart from blogs contains discussion forums and online newspapers, which play the same role as blogs. However, for simplicity, we will call them all "blogs." According to recent Technorati report (see the "Daily posting volume" figure in Ref [9]) the number of daily blog postings did not change during the last year. We thus, for simplicity, assume that $N_l$ is constant in time.

Every day bloggers look for links to blog. These links are of two types. The links of the first type are the links the bloggers copy from recent posts in another blogs (temporary referrers). We assume that the fraction of such *copied* links is $1-\alpha$. Such links are the easiest to find. However, bloggers want to be original, to find some new links, which they did not see in other blogs. For this purpose, they use search engines, web-directories and online encyclopedias (constant referrers). We assume that the fraction of such *found* links is $\alpha$.

As we already did in the single-page model, we do time-discretization with the unit of one day. We assume that the bloggers browse today's entries in the blogosphere and randomly copy from them $(1-\alpha)N_l$ links. To accommodate our time-discrete model we assume that if a blogger found the link today – he will blog it tomorrow. This way for each today's link to the webpage in question tomorrow in blogosphere there will be, on average,

$$\lambda = 1-\alpha \qquad (1)$$

links. As long as $\alpha > 0$, we have a subcritical branching process and, thus, each avalanche of blogging is doomed to end.

The above model contains an unrealistic assumption that the webpages do not differ in their ability to attract bloggers. We have to revise it. When a blogger searches for what to blog today, he browses through other blogs for *interesting* links. He evaluates each link for inclusion in his blog. Each page has certain probability for this decision to be positive. We call this probability page's fitness and denote it as $\varphi$. When the blogger had selected enough links to blog, he stops. Due to its definition as a probability, $\varphi$ is bound between 0 and 1. Obviously, the average probability of positive decision on blogging a considered link is equal to the average fitness of a blogged link. It is different from the average fitness of all WWW pages, because the fit pages are blogged more often. We will denote it as $\langle\varphi\rangle_b$. To collect $(1-\alpha)N_l$ links we need to make $(1-\alpha)N_l/\langle\varphi\rangle_b$ considerations. Thus, each today's blogosphere link to a page with fitness $\varphi$ on average generates

$$\lambda(\varphi) = (1-\alpha)\varphi/\langle\varphi\rangle_b \qquad (2)$$

tomorrow's links. In contrast with the model without fitness, where $\lambda$, given by Eq.(1), is less then 1 for all links, now the links with fitness

$$\varphi > \langle\varphi\rangle_b/(1-\alpha) \qquad (3)$$

are supercritical. To move further we need to compute $\langle\varphi\rangle_b$.

Let us assume that the fitness distribution of the found links is constant in time and denote it $p_f(\varphi)$. The fitness distribution of blogged links on the $n$th day we denote $p_b^n(\varphi)$. The blogosphere on the day n+1 consists of found links and of links copied from the $n$th day entries. The fitness distribution of blogged links on the day n+1 is:

$$p_b^{n+1}(\varphi) = \alpha\frac{\varphi\times p_f(\varphi)}{\langle\varphi\rangle_f} + (1-\alpha)\frac{\varphi\times p_b^n(\varphi)}{\langle\varphi\rangle_b^n} \qquad (4)$$

We can compute the asymptotic distribution by replacing $p_b^n(\varphi)$ and $p_b^{n+1}(\varphi)$ in Eq.(4) with $p_b(\varphi)$. After solving the resulting equation we get:

$$p_b(\varphi) = \frac{\alpha\times\varphi\times p_f(\varphi)/\langle\varphi\rangle_f}{1-(1-\alpha)\varphi/\langle\varphi\rangle_b}. \qquad (5)$$

The obvious self-consistency condition is

$$\int_0^1 p_b(\varphi)d\varphi = 1. \qquad (6)$$

When we know $p_f(\varphi)$ and $\alpha$, $p_b(\varphi)$ in Eq. (5) depends on one unknown parameter $\langle\varphi\rangle_b$. In such case, Eq. (6) can be used to find $\langle\varphi\rangle_b$ and, therefore, $p_b(\varphi)$.

Let us consider a uniform distribution $p_f(\varphi) = 1$. After substituting Eq. (5) into Eq. (6) and performing integration we get

$$\frac{2\alpha\left(-\ln\left(1-(1-\alpha)/\langle\varphi\rangle_b\right)-(1-\alpha)/\langle\varphi\rangle_b\right)}{\left((1-\alpha)/\langle\varphi\rangle_b\right)^2} = 1 \qquad (7)$$

When $\alpha$ is small, for the equality to hold the logarithm must be large and therefore $(1-\alpha)/\langle\varphi\rangle_b$ must be very close to 1. We can replace it with 1 everywhere, but in the logarithm. After this replacement Eq.(7) reduces to:

$$(1-\alpha)/\langle\varphi\rangle_b = 1 - \exp\left(-1/(2\alpha)-1\right) \qquad (8)$$

Thus for small $\alpha$ the factor $(1-\alpha)/\langle\varphi\rangle_b$ is only slightly less than 1. By substituting Eq.(8) into Eq.(2) we get $\lambda(\varphi) = \left(1-\exp(-1/(2\alpha)-1)\right)\varphi$. This means that $\lambda(\varphi)$ is less than 1 for all values of $\varphi$ and the branching process is subcritical. However, for the fittest links it is very close to critical. For example, when $\alpha = 0.1$, we get $\lambda(1) \cong 0.9975$. When we match this model with the single-page model, we see that $rN = \lambda(\varphi) \approx \varphi$. Therefore, our simulation parameters correspond to the value of the fitness of $\varphi = 0.95$.

Let us now consider the case $p_f(\varphi) = 2 - 2\phi$. After substituting Eq. (5) into Eq. (6) and performing integration we get

$$\frac{6\alpha\left(\left(1-\frac{1-\alpha}{\langle\varphi\rangle_b}\right)\ln\left(1-\frac{1-\alpha}{\langle\varphi\rangle_b}\right)+\frac{1-\alpha}{\langle\varphi\rangle_b}-\frac{1}{2}\left(\frac{1-\alpha}{\langle\varphi\rangle_b}\right)^2\right)}{\left((1-\alpha)/\langle\varphi\rangle_b\right)^3} = 1 \qquad (9)$$

Similar to Eq.7, Eq.9 is a transcendental equation for $\langle\varphi\rangle_b$. Obviously, $\langle\varphi\rangle_b$ is bounded between $1-\alpha$ and 1. The upper bound is in place because the average cannot exceed the maximum value of a variable. The lower bound comes because below it the argument of the logarithm in Eq.(9) becomes negative (or, alternatively, $p_b(\varphi)$, given by Eq. (5), becomes negative for $\varphi > \langle\varphi\rangle_b/(1-\alpha)$ ). It is easy to see that the function $p_b(\varphi)$, given by Eq.(5), increases for all values of $\varphi$ when $\langle\varphi\rangle_b$ decreases. Thus, the minimum value of $p_b(\varphi)$ for all values of $\varphi$, and, consequently, of the integral $\int_0^1 p_b(\varphi)d\varphi$ (which is also the R.H.S. of Eq. 9) is reached when $\langle\varphi\rangle_b = 1$. After substituting $\langle\varphi\rangle_b = 1$ into the R.H.S. of Eq. 9, we get that its minimum value is $6\alpha\left(\alpha\ln(\alpha)+1-\alpha-(1-\alpha)^2/2\right)/(1-\alpha)^3$.
One can show that the above expression is always less than 1 when $0<\alpha<1$. Similarly by substituting $\langle\varphi\rangle_b = 1-\alpha$ into the R.H.S. of Eq. 9, we get its maximum value: $3\alpha$. Thus when $\alpha \ge 1/3$ we can always find a value of $\langle\varphi\rangle_b$ which satisfies Eq. 9. However, when $\alpha < 1/3$ Eq.(9) can not be satisfied by any choice of $\langle\varphi\rangle_b$. Remember, however, that when we derived Eq.(5) from Eq.(4) we performed a division by $1-(1-\alpha)\varphi/\langle\varphi\rangle_b$, which, in the case $\langle\varphi\rangle_b = 1-\alpha$, is zero for $\varphi = 1$. Thus, Eq.(5) is correct for all values of $\varphi$, except for 1. In the case when $\langle\varphi\rangle_b = 1-\alpha$ and $p_f(\varphi) = 2-2\phi$ Eq.(5) (which is correct for $\varphi \ne 1$) gives $p_b(\varphi) = 6\alpha\varphi$. We can satisfy Eq.(6) by choosing

$$p_b(\varphi) = 6\alpha\varphi + (1-3\alpha)\delta(1-\varphi). \qquad (10)$$

As $\langle\varphi\rangle_b = 1-\alpha$, Eq.(2) gives $\lambda(\varphi) = \varphi$. This means that the links with the maximum fitness $\varphi = 1$ are exactly critical, while the rest of the links are subcritical.

One can consider a more general fitness distribution $p_f(\varphi) = (\theta+1)(1-\varphi)^\theta$. In the cases $\theta = 0$ and $\theta = 1$ we recover the two just discussed distributions. One can show that when $\theta < (2\times\alpha)/(1-\alpha)$ the asymptotic form of $p_b(\varphi)$ is similar to the case of $\theta = 0$. In the opposite case it is similar to the case of $\theta = 1$, that is it has a delta function part. One can also show that any $p_f(\varphi)$ distribution, which is finite at $\varphi = 1$, is similar to the uniform distribution in this respect. We do not know what the actual distribution of fitness is, but we see that with a wide class of links' fitness distributions the blogosphere self-organizes into a critical state.

Applications of the model are not limited to the web. Recently we used this model to develop a mathematical theory of citing [8]. The only difference is that, instead of bloggers copying links from other blogs, scientists are copying citations from other papers. One can also apply this model to the dynamics of spreading of other elements of culture like books, films and fashions.